\documentclass[%
twocolumn,
superscriptaddress,
 amsmath,amssymb,
 aps,prl
]{revtex4}

\usepackage{amsmath}
\usepackage{amssymb}
\usepackage{wasysym}											%
\usepackage{graphicx}
\usepackage{natbib}												%
\usepackage{fancyhdr}											%
\usepackage{mathtools}										%
\usepackage[usenames,dvipsnames]{xcolor}  %
\usepackage[utf8]{inputenc}               %
\usepackage{multirow}                     %
\usepackage{makecell}                     %
\usepackage[overload]{empheq}             %
\usepackage{ulem}

\definecolor{amaranth}{rgb}{0.9, 0.17, 0.31}
\begin{document}

\title{Magnetic helicoidal dichroism in reflection by magnetic vortices}

\author{Thierry Ruchon}
\email{thierry.ruchon@cea.fr}
\affiliation{Universit\'e Paris-Saclay, CEA, CNRS, LIDYL, 91191 Gif-sur-Yvette, France}

\author{Mauro Fanciulli}
\email{mauro.fanciulli@u-cergy.fr}
\affiliation{Universit\'e Paris-Saclay, CEA, CNRS, LIDYL, 91191 Gif-sur-Yvette, France}
\affiliation{Laboratoire de Physique des Mat\'eriaux et Surfaces, CY Cergy Paris Universit\'e, 95031 Cergy-Pontoise, France}

\begin{abstract}
Studying magnetization configurations of ever more complex magnetic structures has become a major challenge in the past decade, especially at ultrashort timescales. Most of current approaches are based on the analysis of polarization and magnetization-dependent reflectivity. Based on our joint publication XXX XX XXXXXX, we introduce a different concept, centered on the coupling of magnetic structures with light beams carrying orbital angular momentum (OAM). 
Upon reflection by a magnetic vortex, an incoming beam with a unique value $\ell$ of OAM gets enriched in the neighboring OAM modes $\ell\pm 1$. It results in anisotropic far-field profiles, which leads to a Magnetic Helicoidal Dichroism (MHD) signal. The analysis of MHD allows to retrieve the complex magneto-optical constants with excellent precision. This method, which does not require any polarimetric measurement, is a new promising tool for the identification and analysis of magnetic configurations such as vortices, with a possible extension to the femtosecond to attosecond time resolution.    
\end{abstract}
\maketitle
Magnetic nanostructures play a central role in modern technological applications \cite{Chen2016,Dieny2017}, where prominent examples are data storage \cite{Yamada2007}, data transfer \cite{Choi2014,Locatelli2015,Wei2019}, new computing architectures \cite{Torrejon2017, Romera2018} or  biomedical applications \cite{Leulmi2015,Peixoto2020}. Among a great variety of structures in two \cite{Zhang2020} or three dimensions \cite{FernandezPacheco2017}, magnetic vortices (MVs) are particularly promising \cite{Shinjo2000,Bader2006}. They appear in mesoscopic circular dots much larger than their thickness and consist of a curling in-plane magnetic configuration and an out-of-plane core. Their helicity and polarity, respectively the sense of the magnetic curling and core, allow to describe them as topologically protected quasiparticles that are particularly robust against perturbations \cite{Braun2012}. Furthermore, they can be driven out of equilibrium  by magnetic fields \cite{Choe2004} or spin polarized currents \cite{Yamada2007} with rich sub-nanosecond dynamics \cite{Guslienko2006}, offering a way to manipulate them \cite{Leulmi2015,Fu2018}.

MVs have been intensively studied using several imaging techniques, such as magnetic force microscopy \cite{Shinjo2000}, Lorentz microscopy \cite{Schneider2000} or spin-polarized scanning tunneling microscopy (STM) \cite{Wachowiak2002}. These techniques, while having atomic level spatial resolution, are restricted to slow dynamics; their extension below microsecond resolution remains challenging \cite{Houselt2010,Tian2018}. Alternatively, at the price of lower spatial resolution, optical methods using ultrashort laser pulses give access to femtosecond dynamics \cite{BeaurepairePRL1996, Kirilyuk2010,siegrist2019}. They exploit magneto-optical effects, such as the magneto-optical Kerr effect in reflexion (MOKE), or Magnetic Circular or Linear Dichroism in transmission (MCD, MLD), which are all consequences of light's polarization and surface's magnetization dependence of the complex optical indices. The spatial resolution is limited by the focal spot size, ultimately related to the wavelength of light.  Using X-rays, it was possible to combine tens of nanometer spatial resolution with picosecond time resolution using PhotoEmission Electron Spectroscopy \cite{Choe2004} or STM combined with MCD and MLD \cite{Vansteenkiste2009}. Furthermore, analysis of X-Ray scattering patterns allows the statistical determination of average magnetic structures of $100$~nm size \cite{Chauleau2018} with femtosecond resolution \cite{Spezzani2014}. This latter approach, which considers an incoherent scattering of the incoming light, was also proposed to probe MVs \cite{Veenendaal2015}. Conversely, for imaging the exact magnetic structures, the coherence of High Harmonic Generation (HHG) sources was lately exploited in combination with MCD in a coherent diffraction imaging (CDI) setup, yielding images with $50$~nm resolution \cite{Kfir2017}. Here, the image retrieval relies on the analysis of the dichroic diffraction patterns for beams of opposite circular polarization, corresponding to opposite Spin Angular Momenta (SAM).  

In the joint publication \cite{Fanciulli} we investigate a complementary approach exploiting the Orbital Angular Momentum (OAM) of light, which is indexed by an integer $\ell \in \mathbb{Z}$. We show that spatially inhomogeneous magnetic structures yield a so-called Magnetic Helicoidal Dichroism (MHD) for beams of opposite OAM. Both MCD and MHD are linked to the magneto-optical constants. However, MHD primarily depends on the symmetry of the magnetic structure through its azimuthal mode content, making it extremely promising to identify structures, including their signs. It can be observed with any polarization of coherent light beams and vanishes for uniform magnetization.
MVs, which have a very simple decomposition on the azimuthal modes, are privileged test cases for MHD, yielding very simple expressions, and an amplitude in the $10\%$ range.

In this Letter, we theoretically describe MHD for MVs. First we analyze the mode content of a light beam reflected by a MV, showing a selective population of OAM modes. Then we analyse the corresponding far-field images, showing the appearance of MHD. Finally, we show how MHD can be linked to the magneto-optical constants, thus offering a new way to access the magnetic properties of MVs and other magnetic structures.\\

%
\begin{figure}[!htp]
\centering
\includegraphics[width=0.5\textwidth]{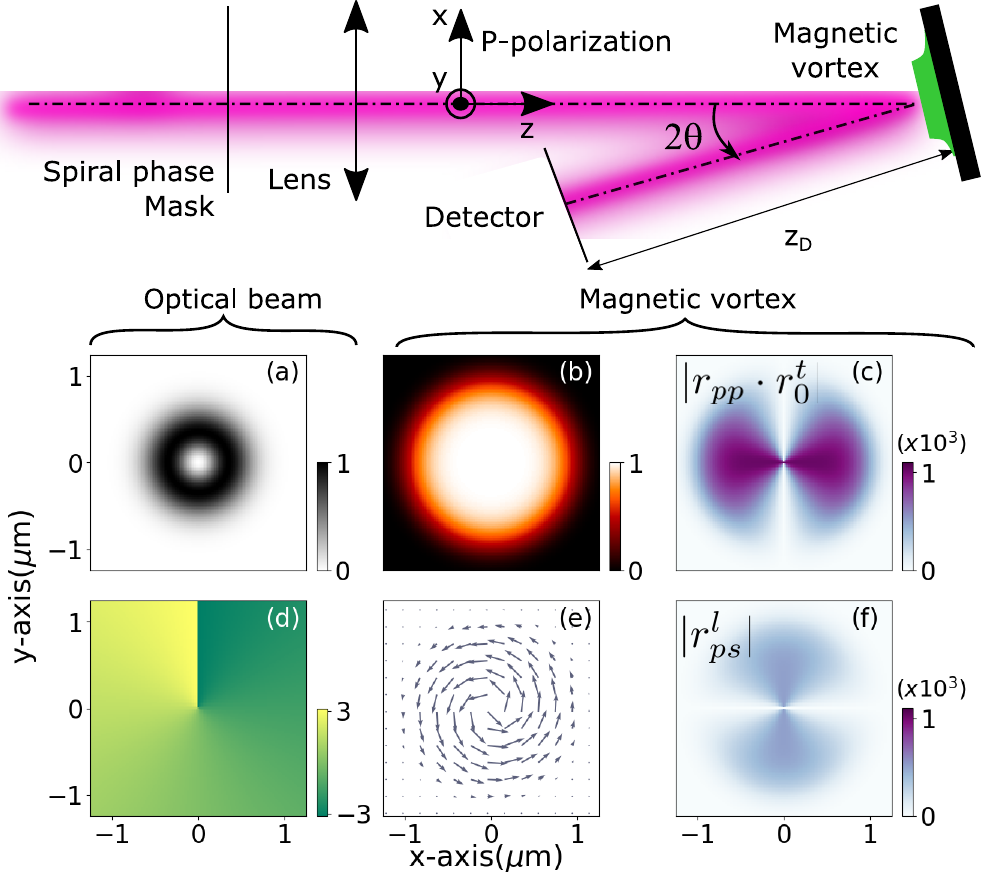}
\caption{
(top) Layout of the physical situation. (bottom) Intensity (a) and phase (d) maps of the incoming beam. Amplitude (b) and direction (e) of the magnetization vector of the MV for $m=+1$. (c) Map of the amplitude of the MOKE constants $|r_{pp}\cdot r_0^t|$ and (f) $|r_{ps}^l|$. \label{Fig1}}
\end{figure}
\begin{figure}[!htp]
\centering
\includegraphics[width=0.5\textwidth]{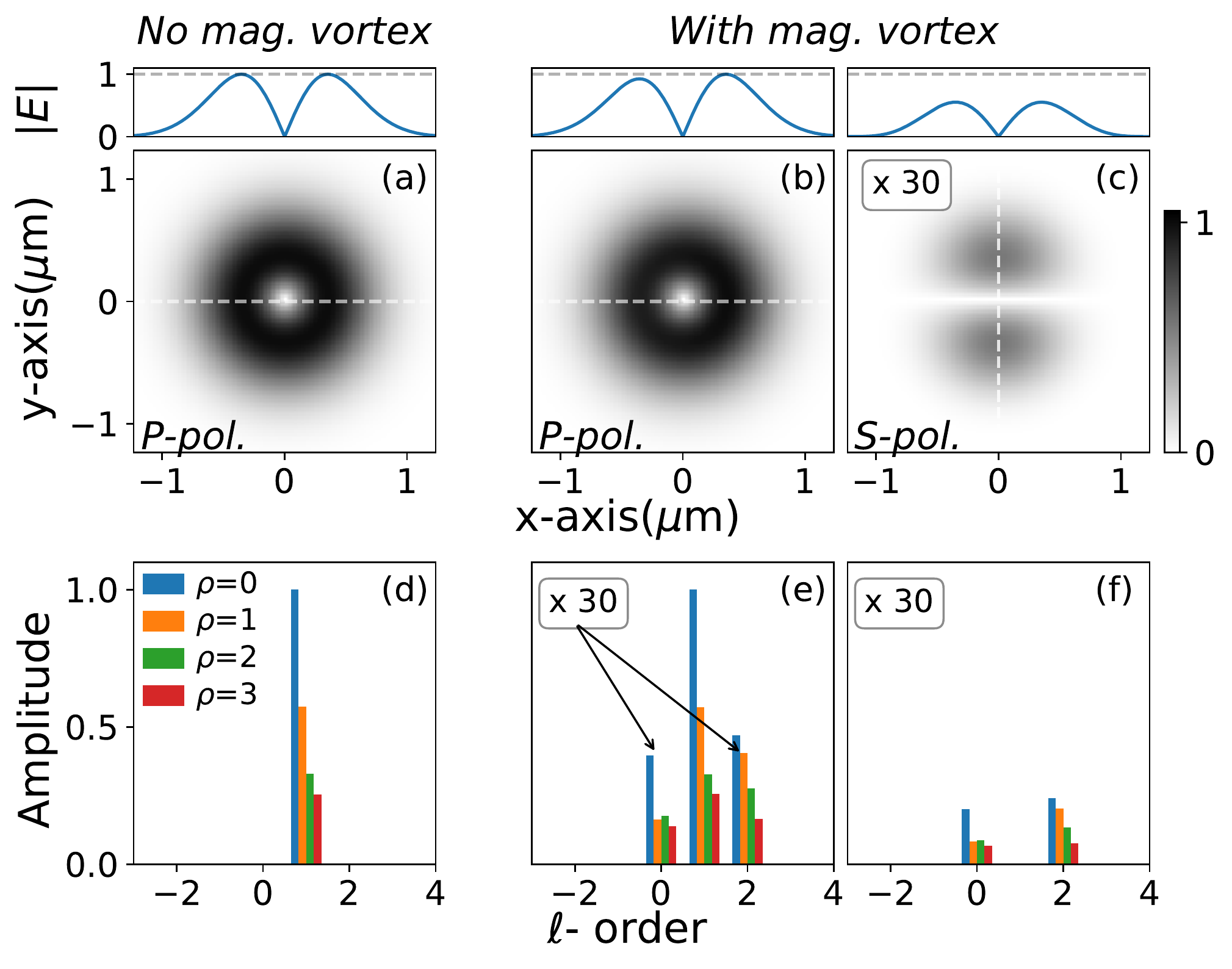}
\caption{
(a) Electric field amplitude after reflection of a P-polarized beam with OAM $\ell=1$ by a magnetic dot with constant magnetization direction and (b,c) by a MV considering P and S outgoing polarizations respectively. The profile along the white dashed lines are shown in the top panels. (d-f) Magnitude of the coefficients of the decomposition on the LG basis $(\ell,\rho)$.\label{Fig2}}
 \end{figure}
\begin{figure}[!htp]
\centering
\includegraphics[width=.5\textwidth]{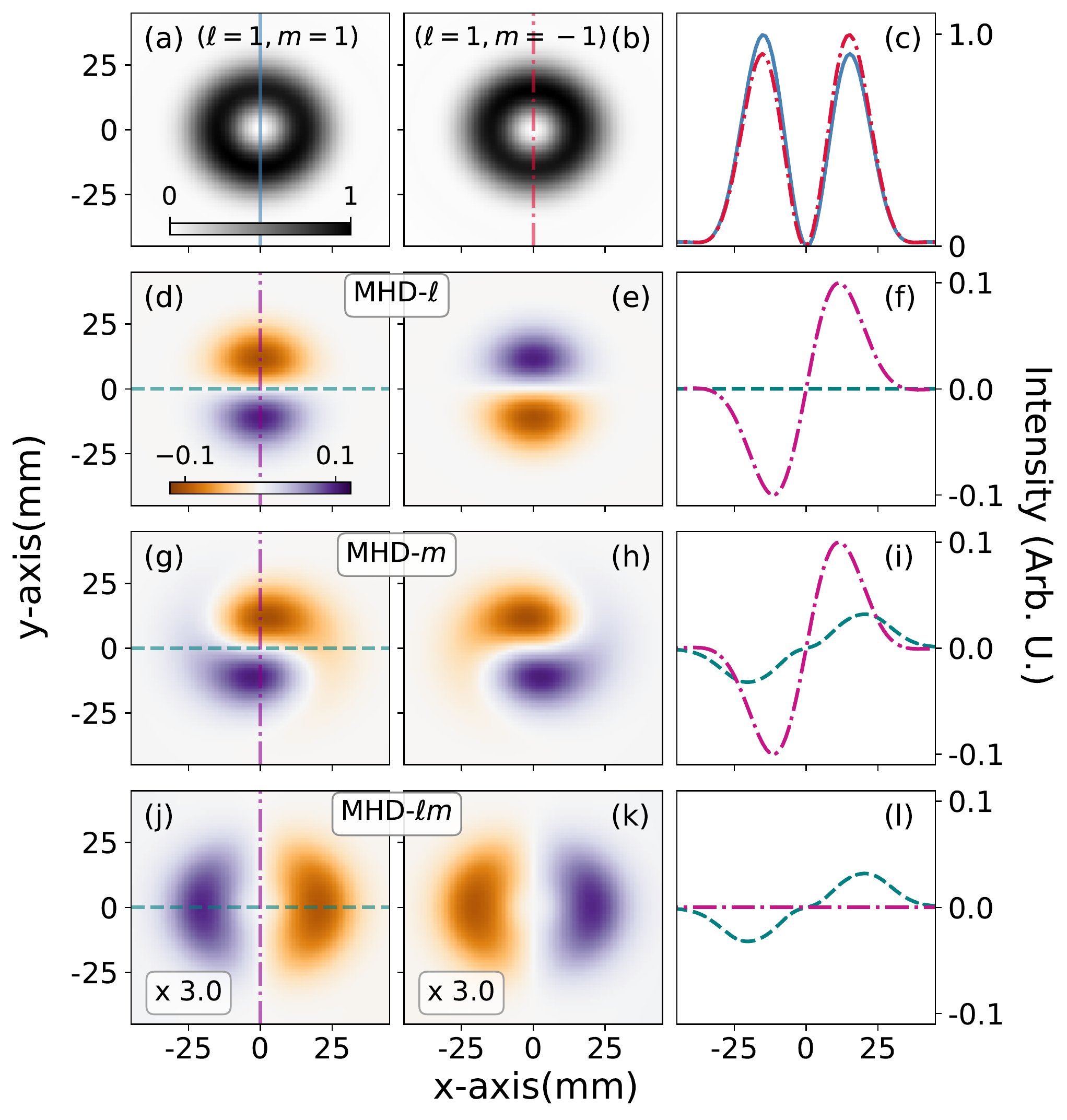}
\caption{
(a) Far-field intensity for $(\ell,m)=(1,1)$ and (b) for $(1,-1)$, with same color scale. (c) Profiles through the $y$ axis for map (a) (blue straight) and (b) (red dotted). (d-e):  MHD-$\ell$, (g-h):  MHD-$m$, (j-k):  MHD-$\ell m$ according to Table~\ref{tbl:symmetries}, and (f,i,l) corresponding lineouts.
\label{FigDichroism}}
\end{figure}
\begin{figure}[!htp]
\centering
\includegraphics[width=.495\textwidth]{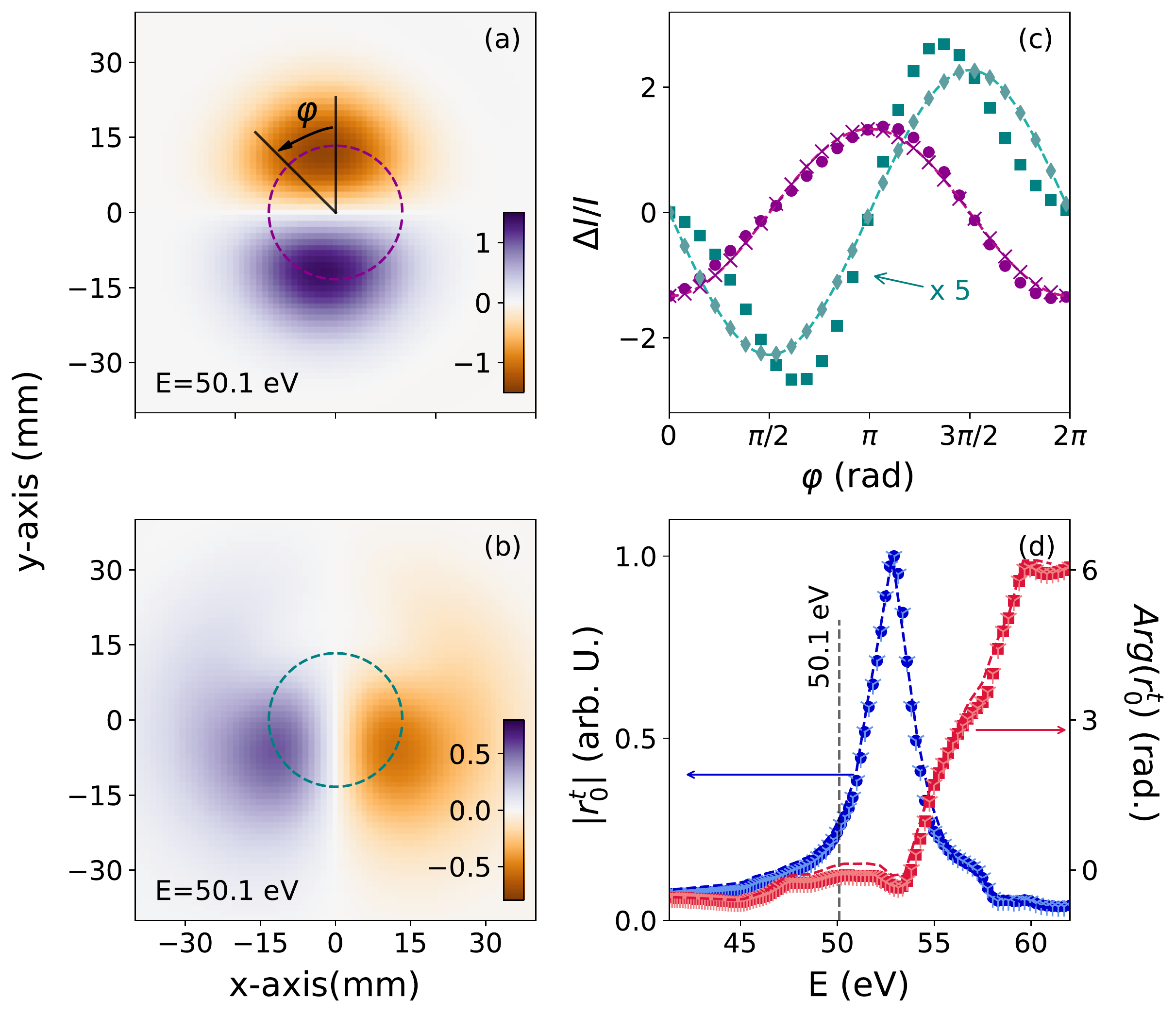}
\caption{
(a) MHD-$\ell$ and (b) MHD-$\ell m$ for $\theta=5^\circ$ and photon energy of $h\nu=50.1$~eV,  to be compared with Fig.~\ref{FigDichroism}(d),(j) respectively, where $\theta=0^\circ$ and $h\nu=52.8$~eV. (c) Lineouts along the circles drawn in panel (a) (purple  dots) and (b) (green  squares), together with the fits from Eq.~\eqref{eq:MHDell}-\eqref{eq:MHDellm}. Purple crosses and green diamonds correspond to the same lineouts for $\theta=0^\circ$. (d) $h\nu$ dependence of amplitude (blue, left axis) and phase (red, right axis) of the retrieved magneto-optical constant $r^t_0$ for $\theta=5^\circ$ (full symbols) and $\theta=0^\circ$ (3-branches crosses). The dashed lines are the corresponding input $r^t_0$ values. Amplitudes are normalized to $1$, with same normalization constant for tilted and not tilted cases. Phases are set equal at resonance.
\label{spectrum}}
 \end{figure}

Our model is sketched in Fig.~\ref{Fig1}.
We consider a $P$ polarized collimated gaussian light beam of wavelength $\lambda$, acquiring OAM $\ell=1$ by going through a spiral phase mask, focused on a MV. Intensity and phase maps of the beam are shown in Fig.~\ref{Fig1}(a,d). The MV is modelled as a ferromagnetic dot, with the sign of its helicity designed by $m=\pm 1$. Amplitude and direction of the magnetization are shown in Fig.~\ref{Fig1}(b,e) for $m=+1$. The MV is tilted by $\theta$ with respect to the $z$-axis of the incoming beam. Although the conclusions are unchanged for large angles and odd values of the incoming OAM, we will first consider $\theta=0^\circ$ to highlight the MHD [Figs.~\ref{Fig2}-\ref{FigDichroism}], and then set a small $\theta=5^\circ$ for a realistic experiment [Fig.~\ref{spectrum}], in order to avoid spurious anamorphisms \cite{Fanciulli}.
We also consider the incoming beam perfectly centered on the MV \cite{Fanciulli}. 

We consider wavelengths significantly larger than the rugosity of the dot's surface, expected below the nanometer. The beam waist at focus is smaller than the MV diameter, avoiding the treatment of edge diffraction. These hypothesis justify the coherent approach proposed here. 
With these constraints, a suitable wavelength range for MHD is $10 \text{\,nm}\lesssim \lambda \lesssim 1000$~nm, which covers magnetization sensitive electronic excitations at optical and core transitions in most elements of interest for magnetic materials.

Beyond the regular Fresnel coefficients $r_{pp}$ and $r_{ss}$ for respectively the P ans S-polarized electric fields, MOKE coupling constants $r_0^t$ and $r_{ps}^l$ are considered. $m_l$, $m_t$ are the in-plane magnetization components along the transverse and longitudinal directions with respect to the scattering plane, normalized by the saturation magnetization. The reflectivity matrix reads \cite{Fanciulli, piovera2013}:
\begin{equation}
R(r,\phi)=
\begin{pmatrix}
r_{pp}+ r_{pp}\cdot r_0^t \cdot m_t& r_{ps}^l \cdot m_l \\
- r_{ps}^l\cdot m_l & r_{ss}
\end{pmatrix}
\label{eq:reflectionMatrix}
\end{equation}
Due to the curling magnetization in the MV, $R$ depends on the azimuth $\phi$, and also on the radial coordinate $r$ because of the finite dimension of the MV. 
We develop a numerical example using the magneto-optical constants of Fe in the XUV range corresponding to the $3p\rightarrow3d$ electron excitation.
Thus we build a model of the reflectivity coefficients yielding the maps displayed in Fig.~\ref{Fig1}(c),(f). 
We computed the following values for their maxima at $\theta=5^\circ$ \cite{Valencia2006}: $r_{pp}=0.027 e^{-1.38j}$, $r_0^t=0.038e^{-0.11j}$ and $r_{ps}^l=0.00051 e^{-1.49j}$. The two coefficients $(r_{pp}\cdot r_0^t)$ and $r_{ps}^l$ have similar amplitude, respectively 3.8\% and 1.9\% of $r_{pp}$.

The analysis of the beam in the near field after reflection is presented in Fig.~\ref{Fig2}. We compare amplitude (a-c) and Laguerre-Gaussian (LG) modes decomposition (d-f) for two cases: a dot with a single domain of constant magnetization along the $y$ axis and the MV sketched in Fig.~\ref{Fig1}, with projection on the P and S field components. In the first case the beam maintains its symmetry. On the contrary, the MV leads to an asymmetry in the intensity profile of the P-component, and the appearance of an S-component, as apparent in the line profiles displayed on top. 
To better identify these asymmetries, we decompose the computed field on an LG-basis with azimuthal and radial indices $(\ell,\rho)$. Due to the finite size of the MV and the imperfect transformation towards a LG mode by the optical setup, even for the single domain case [Fig.~\ref{Fig1}(g)] several radial $\rho$ modes are populated, but no other azimuthal modes than $\ell=1$.
Strikingly, for the MV case we find instead that also the modes $\ell=0,2$ are populated [Fig.~\ref{Fig1}(h-i)]. The result can be generalized to find that the reflected beam presents a change in populated modes due to magneto-optic interaction corresponding to $\Delta \ell =\pm 1$. 

This simple result is ensured by the particularly favorable case of MV for studying MHD, thanks to its azimuthal symmetry \cite{Fanciulli}. Indeed, the ``selection rule'' can be intuitively retrieved observing that  
the magnetization-dependent terms of the $R$ matrix will have coefficients varying like $\cos \phi$. The azimuthal dependence of the incoming electric field, due to the OAM will read $\cos\left( -\frac{2\pi}{\lambda}z-\ell\phi\right)$. The product of the two will thus show $(\ell\pm 1)\phi$ components, and only these. It is clear from this simple analysis that this ``selection rule'' is valid for any incoming value of the OAM; instead it would fail if the beam was not centered on the MV or for a significant tilt $\theta$. However, for odd incoming values of $\ell$, this latter spurious effect does not mix with the one described here \cite{Fanciulli}. 
Furthermore, we notice that the ratio of the weights of the newly populated azimuthal modes over the incoming one is of the same order as that of magnetic over non magnetic reflectivity constants: $2\%$ for the S-component, $4\%$ for the P-component. 

Since LG modes are eigensolutions of the paraxial propagation equation, the intensity profile in the far field will show interferences of the modes $\ell\pm 1$, resulting in asymmetries. We compute the sum of the intensities of the P and S polarization components for the four different combinations of $(\ell,m)=(\pm1,\pm1)$. Two examples are shown in Fig.~\ref{FigDichroism}(a-b). In the considered configuration, MHD appears only in the $P$ component \cite{Fanciulli}, therefore uncontrolled polarizing mirrors in the XUV beamline will not affect the dichroism.    
In order to evaluate MHD, there are $6$ relevant combinations of differences of $(\ell,m)$ values, as listed in the upper triangular part of Table~\ref{tbl:symmetries}. 
Writing $I_{\ell,m}$ the far field intensity of the reflected beam, we classify these six differences in three kinds of dichroisms:
\begin{subequations}
\begin{align}[left = \empheqlbrace\,]
\text{MHD-}\ell &=I_{\ell,m}-I_{-\ell,m}\\
\text{MHD-}m &=I_{\ell,m}-I_{\ell,-m}\\
\text{MHD-}\ell m &=I_{\ell,m}-I_{-\ell,-m}\,.
\label{eq:MHD}
\end{align}
\end{subequations}
They are shown in Fig.~\ref{FigDichroism}, being non zero in all cases, with values up to $10\%$. Intuitively, this is related to the order of magnitude of the interference term between $\ell=1$ and $\ell=0$ modes: $\sqrt{|r_{pp}|\cdot|r_{pp}\cdot r_0^t|}/(|r_{pp}|+ |r_{pp}\cdot r_0^t|)\simeq \sqrt{|r_0^t|}\simeq 20\%$. 
This interference term makes the dichroism detectable, even if away from the Brewster angle where magneto-optical differential signals are usually enhanced.

The $\phi$ periodicity observed in Fig.~\ref{FigDichroism} is a consequence of the mode content found in Fig.~\ref{Fig2}(e), where the $\Delta \ell =\pm 1$ modes interfere with the fundamental $\ell$ mode. 
The two cases of MHD-$\ell$ for a given $m$ [Fig.~\ref{FigDichroism}(d)-(e)] can be exchanged by time inversion $T$ (i.e. switching the helicity $m$ of the MV), and the same is true for MHD-$\ell m$ for a given initial $m$ [Fig.~\ref{FigDichroism}(j)-(k)]. On the contrary, the two MHD-$m$ cases of a given $\ell$ [Fig.~\ref{FigDichroism}(g)-(h)] are exchanged through parity inversion $P$ (i.e. switching the helicity $\ell$ of the OAM), and corresponds to a truly chiral situation \cite{Barron2009}. Indeed, (g)-(h) are chiral patterns that cannot be superimposed by rotation, while (d)-(e) and (j)-(k) can be exchanged by a $\pi$ rotation.
To summarize, MHD-$\ell$ and MHD-$\ell m$ change sign upon change of initial $m$ or $\ell$ signs, while MHD-$m$ is converted to its mirror image when changing $\ell$.
\begin{table}[h!]
     \begin{center}
     \begin{tabular}{ c}
     \raisebox{-\totalheight}{\includegraphics[width=0.48\textwidth]{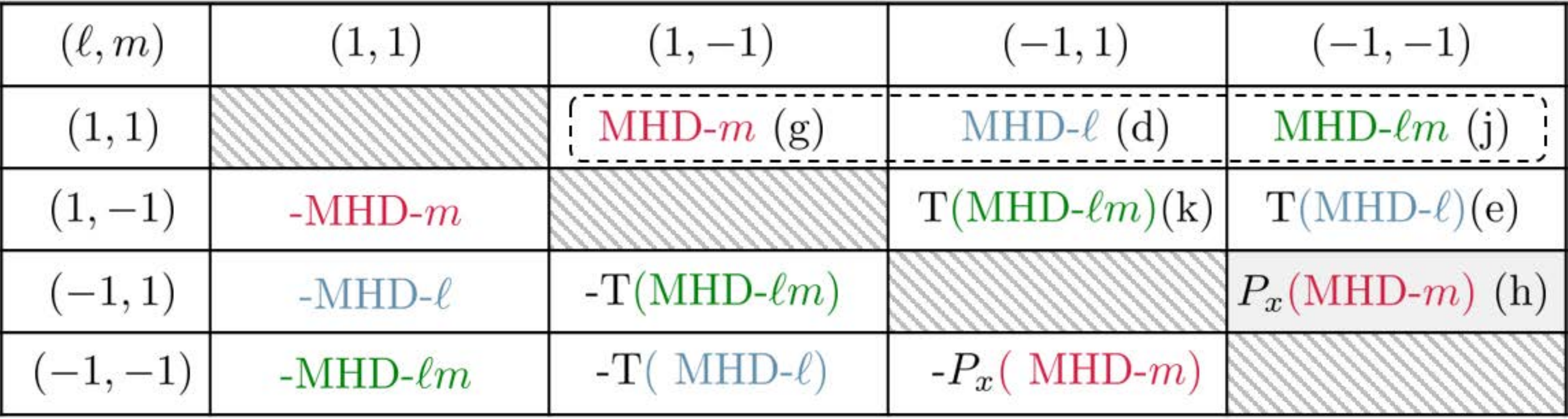}}
      \end{tabular}
\caption{Symmetries of all possible MHD (first column minus first line). They are listed twice (above and below the diagonal), simply corresponding to opposite sign of the difference. Above the diagonal, the dashed rectangle indicates the definition of the three MHD, while the three other configurations are obtained by either a time reversal (T), leading to a sign change, or a parity ($P_x$) reversal, leading to a mirror image. The corresponding panels of Fig.~\ref{FigDichroism} are indicated.}
      \label{tbl:symmetries}
      \end{center}
\end{table} 
It may also be noticed that the information contained in the three MHD is redundant. Indeed, we have $\text{MHD-}\ell=\text{MHD-}\ell m +\text{MHD-} m$ \footnote{This can be obtained as $\text{MHD-}\ell-\text{MHD-}\ell m =-I_{-\ell,m}+I_{-\ell,-m}=-P_x(I_{\ell,m}-I_{\ell,-m})=\text{MHD-} m$, where the parity inversion $P$ acts on the $x$ coordinate.}, so only two measurements are required.

Using the general expressions for MHD found in Ref.~\cite{Fanciulli}, for a saturated MV we have: 
\begin{subequations}
\begin{align}
\text{MHD-}\ell&\approx m|r_0^t|\cos\varphi_0^t \left(-\mathcal{H}_{-1}-\mathcal{H}_{1}\right)\sin\phi\label{eq:MHDell}\\
\text{MHD-}m&\approx m|r_0^t|\sum_{n=\pm 1} \mathcal{H}_{n}\sin(\phi+n\varphi_0^t)\label{eq:MHDm}\\
\text{MHD-}\ell m&\approx m|r_0^t|\sin\varphi_0^t\left(\mathcal{H}_{-1}-\mathcal{H}_{1}\right)\cos\phi,
\label{eq:MHDellm}
\end{align}
\end{subequations}
where 
$\varphi_0^t=\arg\left(r_0^t\right)$ and the function $\mathcal{H}_{n}=\mathcal{H}_{n}(kr,z_D)$ \footnote{The $\mathcal{H}_{n}$ function is given by $\mathcal{H}_{n}(kr,z)=\frac{H_{n,\ell}(kr,z)}{H_{0,\ell}(kr,z)}$, with the function $H_{n,\ell}$ defined in Ref.~\cite{Fanciulli}. $n=\pm1$ for the MV is the index of decomposition of the magnetic structure in the LG basis, $\ell$ is the OAM of the incoming beam, which we considered equal to $1$ here.} 
depends on the beam wavevector $k$, radial parameter $r$ and observation distance $z_D$. 
Interestingly, we observe that two MHD signals, MHD-$\ell$ and MHD-$\ell m$ for example, allow to extract the complex magneto-optical constant $r_0^t$ by fitting the intensity maps to $\sin\phi$ and $\cos\phi$ functions. 

To test this prediction, we compute MHD-$\ell$ and MHD-$\ell m$ maps for several wavelenghts, now with $\theta\neq0^\circ$ for a realistic experiment. They are shown in Fig.~\ref{spectrum}(a),(b) respectively, for $\lambda$ off resonance. 
The tilt induces an extra dissymetry, but with a $2\phi$ symmetry which does not interfere with the MHD $\phi$ symmetry \cite{Fanciulli}. Indeed the two MHD maps, especially MHD-$\ell m$, are modified compared to those of Fig.~\ref{FigDichroism}(d),(j) where we had $\theta=0^\circ$. The fit of the lineout taken along the shown circles is excellent for MHD-$\ell$, while poor for  MHD-$\ell m$ [Fig.~\ref{spectrum}(c)]. However, since tilt and MHD have different symmetry, the fitted cosine still corresponds to the same lineout extracted from the $\theta=0^\circ$ case. Upon normalization of amplitude and phase, we can retrieve the complex $r_0^t$, taking care of scaling the circle with $\lambda$. The comparison to their initial values plugged in the model is excellent[Fig.~\ref{spectrum}(d)], whether the tilt is taken into account or not. Also, the choice of the fitting circle radius has no influence, as long as it gives an intense signal for all wavelengths.\\

In conclusion, we presented an analytical and numerical model of a beam carrying OAM reflected by a MV. Because of magneto-optic interaction, the incoming $\ell$ mode is redistributed into the $\ell\pm1$ modes. Consequently, the far field intensity of the reflected beam is spatially asymmetric because of interference of different modes, and results in a dichroism signal when switching the helicity of the VB (MHD-$\ell$) or of the MV (MHD-$m$). The two are qualitatively different, so also switching both leads to dichroism (MHD-$\ell m$). As an application, we showed how to use MHD in order to extract the value of a MOKE constant with high sensitivity without any polarization device in the experiment. Furthermore, thanks to the complete model presented in the joint publication \cite{Fanciulli}, it is straightforward to extend this approach to other targets, from antivortices to virtually any inhomogeneus magnetic structure, with the possibility to tailor the most suitable experimental conditions in terms of light polarization and reflection geometry. Conversely, different magnetic structures could be engineered in order to analyze the OAM content of a light beam. 

MHD is both a good platform for the basic study of light-matter interaction and a potentially rich spectroscopic tool. On one side, we have the coupling of a beam carrying topological charge with a magnetic material, with MV being a particularly interesting case given its topological nature as well. Also, we did not consider the microscopic, local differential response when the beam is reflected, and coupling with SAM \cite{Kfir2017} would certainly enrich the method. On the other side, since the XUV spectral range is accessible to HHG sources and to free electron lasers, a natural extension to the study of ultrafast dynamics in the femtosecond and attosecond regimes is conceivable. This would provide an access point to the dynamics of MV, which are known to respond to femtosecond pulses \cite{Fu2018}, and potentially to manipulate them. Indeed, a possible effect on the MV such as moving, twisting or switching was not considered, and would require further investigations.

\section{Acknowledgments}
We are grateful to  Maurizio Sacchi and Giovanni de Ninno for stimulating discussions and bringing this subject to our attention. Financial support from the Agence Nationale pour la Recherche (under Contracts No. ANR11-EQPX0005-ATTOLAB and No. ANR14-CE320010-Xstase), and by the Swiss National Science Foundation project No. P2ELP2\_181877 are acknowledged.


\bibliographystyle{unsrt}
\end{document}